\documentclass[10pt]{amsart}

\usepackage{amssymb}
\usepackage{color}

\newcommand{\tr}{\operatorname{tr}}

\newtheorem{theorem}{Theorem}[section]

\theoremstyle{definition}

\newtheorem{problem}[theorem]{Problem}

\theoremstyle{remark}

\numberwithin{equation}{section}

\begin{document}
\bibliographystyle{amsplain}
\title[Hamiltonian complexity]{Hamiltonian complexity}

\author{Tobias J. Osborne}
\email{tobias.osborne@rhul.ac.uk}


\date{\today}

\begin{abstract}
In recent years we've seen the birth of a new field known as \emph{hamiltonian complexity} lying at the crossroads between computer science and theoretical physics. Hamiltonian complexity is directly concerned with the question: \emph{how hard is it to simulate a physical system?} Here I review the foundational results, guiding problems, and future directions of this emergent field.
\end{abstract}

\maketitle

\section{Introduction}
In the natural sciences complex behaviour is usually exhibited by large numbers of interacting particles moving according to simple underlying physical laws. \emph{Many body theory} is the area of physics aimed at explaining these systems, a task often requiring the study of equations comprising large numbers of variables. Problems involving many variables like these are a recurring theme throughout the natural sciences and there are several approaches to understanding their properties. The program that concerns us here is that of \emph{computational complexity theory}, which is the branch of computer science and mathematics focussed on the task of classifying their \emph{computational difficulty}. 

There are striking parallels between the central objects of study in many body theory and computational complexity theory. For example, the \emph{hamiltonians} of many body physics bear some superficial resemblance to the \emph{propositional formulae} studied in complexity theory. It was quickly realised that this analogy runs deeper: the equilibrium states of many physical systems directly correspond to satisfying assignments of variables in certain families of propositional formulae. Following this it made sense to apply the apparatus of computational complexity theory, aimed at answering the question ``how hard is it to decide a propositional formula?'', to the study of many body physical systems. Although these analogies were explored in some detail during the 80s and 90s it was only with the advent of quantum information theory that we have witnessed an intensified cross-fertilisation between computational complexity theory and many body physics, leading to the birth of \emph{hamiltonian complexity}. 

The central objective of hamiltonian complexity is thus to answer the question: ``\emph{how hard is it to simulate a physical system}?" Computational complexity theory provides us with the apparatus to quantify what is meant by ``how hard'' and many body theory, in particular, condensed matter physics, provides us with a rich family of strongly correlated systems to study.

Hamiltonian complexity is interesting precisely because there are families of systems that could be, in principle, engineered and whose physical properties are hard to simulate. The canonical example here is due to Barahona  \cite{barahona:1982} who described an Ising model in a nonuniform magnetic field whose \emph{ground-state energy} is hard to calculate. By ``hard'' we mean that the computational resources required to estimate it grow too quickly with the system size, so that no  computer can carry out the task in a reasonable time. Further, the origin of this hardness isn't just a lack of imagination: no matter how clever you are there is no way to efficiently extract predictions about these systems\footnote{This is because merely the \emph{existence} of an efficient method to compute the ground-state energy of Barahona's model would serve as a counterexample to the $\mathsf{P}\not=\mathsf{NP}$ conjecture. I'll say more about this later.  Obviously, for this review, I am adopting the position that this conjecture is true. If it were false then this would have some spectacular consequences for the natural sciences...}. This peculiar situation is initially disturbing: if one cannot solve the model in a reasonable time, even with access to a supercomputer, then how can the model be said to have \emph{predictive} content? Could it be that there exist systems whose behaviour is simply too hard too predict? This is reminiscent of chaos theory, but the origin of the lack of predictability is rather different. 

So far this discussion applies only to classical systems. However, it was realised that \emph{quantum computers} would be capable of simulation tasks exceeding their classical counterparts. Indeed, a quantum computer can simulate the real-time dynamics of naturally occurring many body systems, both bosonic and fermionic \cite{nielsen:2000a}. With this observation came the hope that quantum computers would also be able to simulate the \emph{equilibrium} dynamics of, e.g., Barahona's model, and much effort was expended on the task of developing such a quantum simulation algorithm. Unfortunately -- or fortunately, depending on your point of view -- it became increasingly evident that such quantum algorithms couldn't always succeed  \cite{aaronson:2008a}. Concomitant with these investigations has been a parallel effort at classifying the \emph{quantum complexity} of computational problems, culminating in the construction of quantum models which at first sight appear to be simulatable with a quantum computer, yet whose properties are, in a specific sense I'll describe later, ``even more'' intractable than their classical cousins.

The puzzling observation that there are systems whose equilibrium properties are hard to simulate has a relatively simple resolution: it turns out that the assumption that the system ends up in an equilibrium state modelled by the Gibbs distribution is simply \emph{wrong}. This isn't really surprising, and is familiar in the context of spin glasses \cite{binder:1986a}: when a system has access to a large number of metastable configurations it is overwhelmingly to end up in a metastable state rather than the unique lowest-energy state. Indeed, Barahona's model, and its cousins, exhibit precisely this property. Thus, if you were to engineer a ``difficult''  system such as Barahona's, it would take an extremely long time for it to equilibrate to its ground-state, and any classical or quantum simulation algorithm for its ground-state properties would get stuck. Thus we've learnt that the manifold of states accessible by the natural dynamics of a physical system is a tiny subset of the full configuration space of the system. By understanding this corner of configuration space better we can design better variational classes of wavefunctions to best parametrise naturally occurring states.

The outline of this review is as follows. We begin in \S\ref{sec:compcomp} with a brief discussion of computational complexity. In \S\ref{sec:simulationproblem} we discuss the central question of hamiltonian complexity, \emph{the simulation problem}, and review its origins and motivations. In \S\ref{sec:whatishard} we review a large body of literature focussed on the presentation of physical systems that are \emph{hard} to simulate.  Finally, in \S\ref{sec:whatiseasy}, we review the literature focussed on the converse situation of understanding what physical systems \emph{can} be efficiently simulated.

\section{Complexity theory}\label{sec:compcomp}

In computational complexity (see, e.g., \cite{papadimitriou:1994a} or \cite{arora:2009a} for a textbook treatment) one categorises the difficulty of computational problems by organising them into \emph{complexity classes}. To understand which class a problem belongs to you need to study how many basic computational steps (e.g., addition, subtraction) are required to solve the problem for larger and larger inputs. It is typical that the larger the input to the problem is the more computational steps are required to solve it. For example, consider the problem of sorting a list of $n$ numbers in descending order. As the input size $n$ gets larger it takes, \emph{in the worst case}, no more than $O(n\log(n))$ steps to do this. Another example: consider the problem of multiplying two $n\times n$ matrices. This takes, in the worst case, no more than $O(n^{2+\epsilon})$, $\epsilon < 1$, operations to complete \cite{cohn:2005a}. In computational complexity one doesn't  care so much about the exponent on the $n$, rather, that there \emph{is} some finite exponent $\alpha < \infty$ so that, in the worst case, no more than $O(n^\alpha)$ operations are required to solve the problem. 

The class of all \emph{decision problems} -- i.e., problems with a ``yes'' or ``no'' answer -- solvable using a number of basic computational steps scaling no worse than a \emph{polynomial} in the input size $n$ is called $\mathsf{P}$. The exact polynomial rate needed to solve a problem in $\mathsf{P}$ may differ from computer architecture to architecture (the actual definition of $\mathsf{P}$ doesn't depend exactly on the definition of \emph{computer}), but the fact that the rate is polynomial doesn't. This is an admittedly coarse classification because for all practical purposes a problem requiring $n^{1000}$ steps to solve is intractable for large instances. Interestingly, for many problems in $\mathsf{P}$ the exponent seems to be low.

Another fundamental complexity class is $\mathsf{NP}$ which is, roughly speaking, the class of decision problems whose solution may be \emph{checked} using a polynomial number of basic computational steps. At first sight this innocuous definition doesn't hint at any hidden intractability. However, it turns out that $\mathsf{NP}$ contains extremely complex problems. Firstly, it is worth noting that $\mathsf{P}\subset \mathsf{NP}$ since the solution to any problem in $\mathsf{P}$ can be checked using polynomial steps (just solve it yourself). To convince yourself that $\mathsf{NP}$ might contain more problems than $\mathsf{P}$ consider the following situation. Imagine you have some function $f$ of $n$ bit values $x_1, x_2, \ldots, x_n$, $x_j\in \{0,1\}$, and you want to find its minimum value. (One can easily convert a problem such as minimum finding into a decision problem by solving the related problem: is the minimum of $f$ less that $c$? If you can solve this problem then you can use a binary search to find the minimum value up to some precision.) It is convenient to imagine that the bit values $\mathbf{x} \equiv (x_1, x_2, \ldots, x_n)$ represent the orientations of a collection of $n$ spins and that $f$ represents an \emph{energy function} of $\mathbf{x}$. Now, as long as it is possible to compute $f(\mathbf{x})$ using only a polynomial number of operations, one can see that this problem of finding the minimum is \emph{in} $\mathsf{NP}$. (That is, if someone gave you the configuration which is claimed to achieve a value of $f$ less than some $c$ then it is easy to check that this is the case.) But is it in $\mathsf{P}$?

There is one obvious strategy to find a value of $f$ less than some desired $c$: simply search over the space of all bit strings $\mathbf{x}$. However, this task is prohibitively expensive (it takes $O(2^n)$ steps in the input size $n$). Perhaps there is a better strategy? Unfortunately not, unless $f$ has some special structure. For example, if $f$ is comprised of a sequence of random interactions -- each coupling collections of many spins -- the system will be \emph{frustrated} and $f$ will possess many local minima, ruining any local search heuristic.

The canonical example of such a frustrated $f$ is provided by the 3SAT problem: here $f$ is given by a Boolean expression written as a conjunction (i.e., the ``OR'', denoted $\vee$) of a collection of \emph{clauses} $C_j$, each of which is the disjunction (i.e., the ``AND'', denoted $\wedge$) of three \emph{literals} (i.e., a Boolean variable $x_j$ or its negation $\overline{x}_j$). An example of an allowed clause is, e.g, $C_j(\mathbf{x}) = \overline{x}_{j_1} \vee x_{j_2} \vee x_{j_3}$. There are 8 possible choices for which variables can be subjected to flips and all are considered to be possible. You should think of which variables are actually subjected to flips in each clause, and which variables appear in each clause, as being chosen by an adversary. The function $f$ corresponds to a (classical) hamiltonian for a system of $n$ spins with general 3-spin interactions via
\begin{equation}\label{eq:3SAT}
	f = \bigwedge_{j=1}^m C_j(\mathbf{x}) \longleftrightarrow H(\mathbf{x}) = -\sum_{j=1}^m h_j(\mathbf{x}),
\end{equation}
where each of literals $x_j$ are interpreted as numbers via $\mathrm{TRUE} = 1$ and $\mathrm{FALSE} = 0$ and the interaction terms $h_j$ are found by replacing $\vee$ with the standard product operation, e.g., $h_j(\mathbf{x}) = \overline{x}_{j_1} x_{j_2} x_{j_3} \longleftrightarrow C_j(\mathbf{x}) = \overline{x}_{j_1} \vee x_{j_2} \vee x_{j_3}$. A configuration of zeros and ones achieving the minimum value of $H$ corresponds to a \emph{satisfying} assignment of $f$ where all of the clauses evaluate to TRUE. If zero is regarded as a spin up and one is regarded as a spin down then $H$ can be thought of as the hamiltonian for a (possibly frustrated) spin system.

It is not impossible that there \emph{is} a general-purpose efficient algorithm to find the minimum value of $H$, but this would have fabulous consequences because of a fundamental result in computational complexity theory known as the \emph{Cook-Levin theorem}: the 3SAT problem is \emph{complete} for the class $\mathsf{NP}$. The theorem says that if you had an algorithm which could solve 3SAT using a polynomial number of steps, \emph{in the worst case}, then you could solve \emph{any} other problem in $\mathsf{NP}$ efficiently. Any such problem in $\mathsf{NP}$ with the property that an efficient algorithm for it would yield an efficient algorithm for all of $\mathsf{NP}$ is known as an \emph{$\mathsf{NP}$-complete problem}. A problem is known as \emph{$\mathsf{NP}$-hard} if it as at least as hard as the hardest problems in $\mathsf{NP}$. There is no requirement that such problems be in $\mathsf{NP}$; indeed, they may not even be decision problems. The class $\mathsf{NP}$ is now known to contain a vast number of problems, many of which have considerable practical relevance. However, since no efficient solution algorithm has ever been found, nor are there serious proposals, the possibility of $\mathsf{P} = \mathsf{NP}$ is considered to be extraordinarily unlikely. 

Interestingly, there are some surprising examples of systems that appear, at first sight, to be intractable, yet which admit efficient algorithms for their physical properties. A good case in point is that of a \emph{ferromagnetic} Ising model -- where the spin-spin interactions are always positive -- with arbitrary interaction graph. Thanks to Barahona's example (which pertains to an Ising model with both antiferromagnetic and ferromagnetic interactions), one might be tempted to assume that the equilibrium properties of this model cannot be simulated efficiently. However, there is a clever polynomial-time algorithm which samples configurations according to its Gibbs distribution \cite{jerrum:1993a}, to a required precision. 

With the advent of quantum compution the landscape of computational complexity theory has been fundamentally extended: it is natural to introduce a new class of problems extending the classical class\footnote{Strictly speaking, the class $\mathsf{BQP}$ extends the class $\mathsf{BPP}$, which are the decision problems that can be decided with a bounded probability using a polynomial number of arithmetic operations and coin flips.} $\mathsf{P}$, which are easy for a quantum computer to decide. This natural class is known as $\mathsf{BQP}$ and contains several problems thought to lie outside of $\mathsf{P}$, e.g., factoring. The classification of the computational complexity of problems in the presence of quantum computers is known as \emph{quantum complexity theory}.

There is also a natural class of problems known as $\mathsf{QMA}$, analogous to $\mathsf{NP}$. To understand this generalisation it is convenient to view $\mathsf{NP}$ as a game played between two players: (M) a malevolent Merlin, with infinite computational power; and (A) Arthur, limited to using a standard classical computer with polynomial computational power. In this game Arthur asks Merlin questions, such as ``does $f$ have a satisfying assignment?'' and Merlin accompanies the answer with a \emph{proof} $\mathbf{x}$, which Arthur can use to verify the answer. In the case where $f$ doesn't have a satisfying assignment there is no way Merlin can ever fool Arthur that it does. The class $\mathsf{NP}$ is then the class of all decision problems which can be decided via this game. 

The quantum generalisation of $\mathsf{NP}$ is defined along the same lines, except that Arthur is allowed access to a \emph{quantum computer} with polynomial computational power and Merlin is allowed to send quantum states as a proof. Arthur asks the same questions as before, but now Merlin is allowed to send a quantum state to convince Arthur that a given problem has a ``yes'' answer: Arthur \emph{accepts} that a given problem has a ``yes'' answer if, after running a quantum verification algorithm on Merlin's quantum state, he measures a $1$ and \emph{rejects} if he measures a $0$. We say a decision problem is in $\mathsf{QMA}$ if Arthur always accepts a ``yes'' instance with probability greater than $2/3$ and always rejects a ``no'' instance with probability less than  $1/3$.

A remarkable result of Kitaev (discussed below) establishes a Cook-Levin-like theorem for $\mathsf{QMA}$, and provides a complete problem for it. The main intuition to take away here is that the class $\mathsf{QMA}$ contains ``harder'' problems than $\mathsf{NP}$. 

\section{The simulation problem}\label{sec:simulationproblem}
Typically the way physics has done is as follows: (1) formulate a theoretical model; (2) (approximately) solve the model to make a prediction; then (3) experimentally test the prediction. If the experiment confirms the prediction then declare success. If not, then go back to the drawing board. Each of these three steps typically requires considerable ingenuity, insight, and resourcefulness to carry out. However, of these three steps, the second has a somewhat algorithmic flavour: if one only threw enough computational resources at the model then surely it could be solved? This echoes Hilbert's \emph{Entscheidungsproblem}: perhaps for any physical model, when sufficiently clearly formulated, there exists a procedural algorithm to extract predictions, at least up to some acceptable error margin? The analogy with the Entscheidungsproblem here has been specifically chosen to foreshadow the answer: \emph{no}, there is no such general efficient method. 

Hamiltonian complexity has been developed to answer precisely the question: \emph{how hard is it to simulate a physical system?} To attack this vaguely phrased question it is convenient to introduce a decision problem to capture its essential elements. This problem, called the \emph{simulation problem}, is described as follows. Firstly, we need to choose a physical system. In all cases we assume the system (usually quantum) is modelled with a hamiltonian $H$. A crucial ingredient in the simulation problem is that we don't just have one system but rather the specification of a \emph{family} $\{H_n\}_{n=1}^\infty$ of physical systems (usually indexed by the number, $n$, of particles involved). This is fundamental because in order to answer the question ``how hard is it to simulate $H_n$?'' using the tools of computational complexity we are going to quantify the hardness using the \emph{scaling} $f(n)$ with $n$ of the number  of arithmetic operations required to simulate the system. 

Next, ``simulate'' is far too vague: we also need to specify what state the system is in, i.e., thermal equilibrium or dynamically evolving? Additionally we need to say \emph{what} we wish to simulate because for systems comprising $n$ particles there are typically exponentially many possible quantities  which are, in principle, observable, but most of these are experimentally inaccessible. Thus the final ingredient of the simulation problem is the specification of a family $\{A_n\}_{n=1}^\infty$ of observables to simulate. Finally, it is unlikely that, constrained as we are by the finiteness of machine precision, we can make perfect predictions for $A_n$. Therefore we specify two expectation values $\alpha_1$ and $\alpha_2$ which our predictions \emph{must} decide between. With these qualifications in mind we now write out the simulation problem.

\noindent
\begin{problem}[The simulation problem] \begin{itemize}\end{itemize}
	\noindent
	Input:
	\begin{itemize}
		\item A specification of a family $\{H_n\}_{n=1}^\infty$ of hamiltonians, a specification of an initial state $\rho_{0;n}$, a family of observables $\{A_n\}_{n=1}^\infty$, a (possibly complex) time $t$, two proposed expectation values $\alpha_1 < \alpha_2$, and an integer $n$.
	\end{itemize}
		Output:
	\begin{itemize}
		\item  ``yes'' if 
		\begin{equation}
			\langle A_n \rangle \le \alpha_1,
		\end{equation}
		or 
		\item ``no'' if 
		\begin{equation}
			\langle A_n \rangle \ge \alpha_2,
		\end{equation}
	\end{itemize}
	where 
	\begin{equation}
		\langle A_n \rangle = \tr(\rho_n A_n),
	\end{equation}
	and 
	\begin{equation}
		\rho_n = \frac{(e^{itH_n})^\dag\rho_{0;n} e^{itH_n}}{\tr((e^{itH_n})^\dag\rho_{0;n} e^{itH_n})}.
	\end{equation}
	Note that, because we allow $t$ to be possibly imaginary, e.g., $t=i\beta$, the trace factor in the denominator may be nontrivial. From now on we drop the subscript $n$ on $H$ and $\rho$ and leave the dependence implicit.
\end{problem}

Two special cases of the simulation problem have received considerable attention in the literature, namely, the \emph{real-time} case where $t$ is real, and the \emph{ground-state} case where $\rho_0 = \mathbb{I}/\tr(\mathbb{I})$ and $\rho = \lim_{\beta\rightarrow \infty} e^{-2\beta H}/\tr(e^{-2\beta H})$ which is discussed in the next section.

As we'll see, much effort has been expended on understanding the simulation problem in the case where the observable $A$ is the hamiltonian $H$ of the system itself. I personally think that there is much to be gained by studying more general observables $A$: not much known about the complexity of the simulation problem for situations of immediate physical relevance, e.g., when $A$ is \emph{bulk} or \emph{extensive} quantity such as the magnetisation of a quantum magnet. But surely this should be an easy problem to simulate? After all, thermodynamics tells us how bulk properties should behave. Obviously it all depends on what tolerances $\epsilon = |\alpha_1-\alpha_2|$ you choose: there should be a smooth transition in the complexity of the simulation problem from hard to easy as $\epsilon$ is varied between, say, $\epsilon = O(2^{-n})$ to $\epsilon = O(n)$. Indeed there is some evidence of such a transition: for $\epsilon$ small enough the problem of estimating the magnetisation is essentially equivalent in hardness to the ground-state problem \cite{gottesman:2011a} and for $\epsilon \sim n$, in the case of certain \emph{regular} lattice systems, there is an efficient approximation scheme to estimate the ground eigenvalue \cite{bansal:2009a}. 

Another direction which has received considerably less attention is that of $t$ complex, but not pure imaginary, nor pure real. This problem is of considerable physical importance as it corresponds to the situation where a system is dynamically evolving at some \emph{finite temperature}, which is the generic situation in experiments where it is usually impossible to cool a system to its ground state. Looking further afield, it is also interesting to propose the study of the computational complexity of systems in the presence of \emph{decoherence}. In both the finite-temperature and the decoherent cases there ought to be a transition in the hardness of the simulation problem from hard at low temperature/decoherence rate to easy for high temperature/decoherence rate.
 
\section{What is hard}\label{sec:whatishard}
In recent years we've seen a string of counterintuitive results presenting realistic physical systems whose observable properties are \emph{hard} to estimate. As previously mentioned, a foundational result was presented in a paper of Barahona \cite{barahona:1982} who proved that for the hamiltonian $H$ of the \emph{classical} Ising model in a nonuniform magnetic field the \emph{ground-state} (i.e., $t = i\lim_{\beta\rightarrow\infty} \beta$) simulation problem is $\mathsf{NP}$-hard for the observable $A= H$ with $\alpha_1 = 0$ and $\alpha_2 = 2$. Barahona's construction  -- pertaining to a classical model which, in our quantum context,  simply means a model whose hamiltonian is diagonal in a product basis -- already presents a family of realistic systems whose physical properties are hard to predict: any efficient general purpose algorithm to simulate the ground-state energy of Barahona's Ising model would already be capable of deciding $\mathsf{P} = \mathsf{NP}$, and hence doesn't exist. It is worth commenting here that these results pertain to the \emph{worst case} where the model is engineered adversarily. It is fairly likely that the \emph{average case} complexity, where the interactions are not chosen adversarily, of a physical model will, in general, be quite different from the worst case.

The most important foundational result for hamiltonian complexity in the quantum setting is that of Kitaev \cite{kitaev:2002a} (for a review see \cite{aharonov:2002a}) who showed that the ground-state simulation problem where $A=H$ is a $5$-local hamiltonian (i.e., a hamiltonian for a quantum spin system whose interactions involve no more than $5$ spins at a time) with $\alpha_1 = 0$ and $\alpha_2 = O(n^{-3})$ is \emph{complete} for $\mathsf{QMA}$. Kitaev's proof exploited a clever trick inspired by Feynman generalising an idea exploited in the proof of the Cook-Levin theorem. This construction has been influential in all subsequent works and can be regarded as a truly quantum generalisation of the Cook-Levin theorem. Thus $\mathsf{QMA}$ has the same property as $\mathsf{NP}$: any efficient classical algorithm for the ground-state simulation problem for Kitaev's $H$ would yield an efficient classical algorithm for any other problem in $\mathsf{QMA}$. Since $\mathsf{QMA}$ contains $\mathsf{BQP}$, this would imply that all quantum computations are classically simulable. 

It may be reasonably argued that a $5$-local hamiltonian isn't particularly natural, and this prompted researchers to explore exactly what kinds systems might have ground-state properties which are hard to approximate. Building on Kitaev's construction a sequence of results refining the original  have emerged \cite{kempe:2003c, kempe:2004a, nagaj:2007a, kay:2007a, oliveira:2008a, aharonov:2009a} culminating in the surprising and counterintuitive result \cite{aharonov:2009a} that there exist even \emph{one-dimensional} quantum spin systems whose ground-state energy is hard to simulate within a precision $\epsilon = O(n^{-\alpha})$, $\alpha \ge 3$. This remains true even when the system is translation invariant \cite{gottesman:2009a}. It is worth pausing a moment to appreciate the full implications of this result. In the classical case the one-dimensional ground-state simulation problem is in $\mathsf{P}$ because one can simply evaluate the partition function via a straightforward product of transfer operators. The results \cite{aharonov:2009a} of Gottesman \emph{et.\ al.}\ conclusively establish that no quantum analogue of the standard transfer operator method exists in the quantum setting. 

In the proof described in \cite{gottesman:2009a} it seems necessary that the local spin dimension needs to be rather large (at least more than, say, $8$ or $9$). It is an interesting question to study whether reducing the local spin dimension makes the problem easier. Some evidence this might be the case arises in the case where constraints are chosen randomly (see the discussion below concerning quantum satisfiability). Here it is known \cite{movassagh:2010a} that there is a transition in the \emph{entanglement} structure of the ground state of one-dimensional quantum systems as the local spin dimension changes from $2$ to $4$. The case of $3$-dimensional local spins seems special. Based on this I would be willing to conjecture that the ground-state simulation problem for one-dimensional chains of spin-$1/2$ degrees of freedom is in $\mathsf{P}$. (Some of my colleagues disagree with me that this could be true. Any evidence either way would be quite interesting.)

While initial research focussed on the setting of distinguishable quantum spins, much remains true in the physically more realistic setting of interacting particle systems comprised of bosons or fermions. Along these lines the simulation problem for a family of systems known as \emph{stoquastic}, whose hamiltonian's off-diagonal entries are negative, has been investigated \cite{bravyi:2008a, liu:2007b, jordan:2010a, bravyi:2009a}. Such systems do not exhibit a ``sign problem'' because the coefficients of the ground state are nonnegative. (All bosonic systems in second quantisation with the standard kinetic+potential energy structure exhibit this property.) Similar results hold here: the ground-state simulation problem for stoquastic hamiltonians has been shown to be complete for the class \emph{stoquastic-$\mathsf{MA}$} which is, roughly speaking, a class generalising $\mathsf{NP}$ with a complexity intermediate between two probabilistic classes $\mathsf{MA}$ and $\mathsf{AM}$ extending $\mathsf{NP}$.  Additionally, the original $\mathsf{QMA}$ results developed above have been used to show that ground-state simulation problem is $\mathsf{QMA}$ hard for both fermions \cite{schuch:2009a} and bosons \cite{wei:2010a}.

A very closely related problem (indeed, dual to) to the simulation problem known as the \emph{density operator consistency problem} has also received attention. Here the task is to decide if, given a list of reduced density operators, whether there exists a global pure state consistent with these reductions. This problem is now known \cite{liu:2006a} to be hard for $\mathsf{QMA}$. The fermionic equivalent, the \emph{$N$-representability problem}, of central importance in quantum chemistry, has also been shown to be $\mathsf{QMA}$ hard \cite{liu:2007a}.

Parallel to this there have been several investigations exploring physically motivated specialisations, generalisations, and alternative settings of the original local-hamiltonian results. Notable results along these lines include studies of the complexity of state preparation via cooling \cite{janzing:2003a} of local hamiltonian systems, and the difficulty of the task of simulating the density of states \cite{shi:2009a, brown:2010a} of a local hamiltonian.

There is one special case of the simulation problem for local hamiltonians which is of particular interest, namely, \emph{quantum $k$-satisfiability}: here the task is to decide whether the hamiltonian is \emph{frustration-free} or not, i.e., whether the ground state is \emph{exactly} the ground state of each interaction term, or is far from such a state. This situation corresponds to the simulation problem where $A = H \ge 0$ with $\alpha_1 = 0$ and $\alpha_2 = 1/\text{poly}(n)$, i.e., the task is now to simply decide if a given state is \emph{exactly} a ground state or far from one when given the additional promise that the hamiltonian is comprised of a sum of projection operators. This problem is arguably a more natural quantum generalisation of 3SAT. The same proof applies, as above, for local hamiltonians to show that this problem is hard in exactly the same settings, i.e., for one-dimensional translation-invariant quantum spin systems. The case of frustration-free \emph{stoquastic} hamiltonians \cite{bravyi:2009a} is also rather interesting: here one can simulate the adiabatic evolution of such models classically.

The quantum $k$-SAT problem has turned out to be a fruitful basis for quantum generalisations of the \emph{random satisfiability} problem, where constraints are chosen randomly. It is now understood that so-called random quantum $k$-SAT exhibits a transition \cite{laumann:2010a, laumann:2010b, laumann:2010c, movassagh:2010a, bravyi:2010a}, analogous to its classical counterpart \cite{kirkpatrick:1994a}, from a phase where there are almost always many possible frustration-free ground states to a phase where there are almost no frustration-free ground states.

A setting which has been the subject of considerable conjecture and interest is that of quantum spin systems with \emph{commuting} interactions \cite{bravyi:2005a, aharonov:2011a, schuch:2011a}. One might be forgiven for initially guessing that such systems are completely classical because all interactions can be simultaneously diagonalised. However, a little thought shows that this simultaneous diagonalisation isn't particularly helpful because the basis in which all the interactions are trivial may be highly entangled. Thus, the ground-state simulation problem may well be harder than its classical counterpart. Indeed, some evidence this might be the case has been recently obtained \cite{aharonov:2011a}. Deciding whether the simulation problem for commuting interactions is harder than $\mathsf{NP}$, or in $\mathsf{NP}$, is an intriguing open problem because it is clear that it requires new ideas going beyond the existing techniques.

\subsection{Guiding problem: the quantum $\mathsf{PCP}$ conjecture}
To date, in all of the constructions of quantum systems whose ground-state energy is hard to simulate, the precision $\epsilon = |\alpha_1-\alpha_2|$, or \emph{promise gap}, for which the problem is known to be hard scales as an inverse \emph{polynomial} in the system's size. That is, the problem is known to be hard only when one requires that the energy is very precisely approximated. It is thus natural to ask to what extent the intractability of these systems is tied to this constraint, i.e., perhaps it is easy to decide if the ground-state energy is small or larger than, say, some \emph{constant fraction} of the number $m$ of interaction terms? 

In the classical setting this question has a dramatic and counterintuitive answer: it is also complete for $\mathsf{NP}$. This result, captured by the celebrated \emph{PCP} theorem\footnote{PCP stands for probabilistically checkable proof.} \cite{arora:1998a, arora:1998b, dinur:2007a}, shows that even the task of approximating the ground-state energy to within a constant factor is hard. 

The PCP theorem already establishes that there are physical systems, indeed classical systems, for whom the simulation problem is hard, even when $\epsilon = |\alpha_1-\alpha_2| = cm$, with $c$ some constant. However, it is entirely natural to suppose that that there should be a corresponding quantum equivalent of the PCP theorem which would say that the ground-state simulation problem is $\mathsf{QMA}$-complete for $\epsilon = |\alpha_1-\alpha_2| = cm$, with $c$ some constant \cite{aharonov:2009b, aharonov:2011a}. 

Some progress towards a possible proof of the conjectured \emph{quantum PCP theorem} has been made whereby an important proof element known as \emph{gap amplification} -- exploited in a recent proof \cite{dinur:2007a} -- has been generalised to the quantum setting, where it is now known as the \emph{detectability lemma} \cite{aharonov:2009b}. However, so far, a fully quantum PCP theorem has resisted all attempts at a proof. 

The quantum PCP conjecture is a fascinating guiding problem because it would require the development of many theoretical tools of independent interest. For example, it is considered likely by practitioners that a proof of the quantum PCP conjecture would aid in the design of error-tolerant architectures for \emph{adiabatic} quantum computation. Additionally, proof techniques developed in the pursuit of the quantum PCP conjecture also have direct applications in the study of the local hamiltonians appearing in condensed matter physics. For example the detectability lemma has already found applications in the study of entropy/area laws for quantum spin systems (see below) \cite{aharonov:2010b}. A quantum PCP theorem would mean that there exist certain realistic hamiltonians for which no variational representation of the ground state with a concise classical description could reproduce the ground-state energy better than a constant precision.

At this point it seems that a proof of the quantum PCP conjecture would have an inductive structure whereby a very bad quantum PCP is iteratively improved. However, there is a stumbling block in any proof following this strategy, namely, the design of at least one, possibly very inefficient, quantum PCP. This is summarised by the following problem: show that a $\mathsf{QMA}$-complete simulation problem for $n$ spins can be encoded as the simulation problem for a local hamiltonian involving an \emph{exponential} (or even \emph{doubly exponential}) number $N$ of spins, i.e., $N = 2^n$, or $N = 2^{2^n}$, with $\epsilon = |\alpha_1-\alpha_2| = cN$, $c$ a constant. In other words, if we don't care how many spins we use is it possible to improve the promise gap $\epsilon = |\alpha_1-\alpha_2|$ \emph{at all} while retaining the hardness of the problem? Historically this was one of the steps toward the classical PCP theorem. What I have in mind here is something like a quantum generalisation of the results of H{\aa}stad \cite{hastad:2001a} on inapproximability of certain problems involving, e.g., solutions to linear equations. 

\section{What is easy}\label{sec:whatiseasy}
In the previous section we saw that considerable effort has been expended on the task of constructing plausible physical models whose \emph{ground-state} properties are hard to approximate. In this section I'll take a look at the flip side of the coin, i.e., understanding what physical properties are easy to approximate. To answer this question concretely requires the description of an \emph{algorithm} which is \emph{correct}, \emph{converges}, and has a polynomial \emph{complexity}. The design of such algorithms is truly an art form, but one informed by physical results. 

The simple and striking conclusion we've drawn so far is that the natural dynamics of a locally interacting system doesn't always take the system an equilibrium state modelled by the Gibbs distribution. The  question then arises: what states can a locally interacting system enter under natural dynamics? The key insight to draw on here is that the dynamics of a locally interacting collection of particles cannot explore the entirety of hilbert space in any reasonable length of time as it takes only a \emph{polynomial number}, in $n$, of parameters to completely specify a local hamiltonian, but to specify a state in hilbert space requires an exponential number of parameters. Therefore, in order for a local system to explore hilbert space via its local dynamics, it ought to require at least an exponential time to visit every location. (Here we are thinking of hilbert space as a manifold rather than a linear space, i.e., we are forgetting the linear structure.) This simple counting argument can be improved \cite{poulin:2011a} and it turns out that the strongest result is much better: it takes a \emph{doubly exponential} time for the dynamics of a local hamiltonian to visit an arbitrary state in hilbert space. 

To specify a pure state in the classical situation one requires only a linear (in $n$) number of parameters, namely, the positions of each of the particles. Another way of saying this is that for any pure classical state one can assign a definite pure state to each of the constituent particles. The main difficulty one faces in simulating a many particle quantum system is that it is difficult to describe a generic \emph{pure} state in the hilbert space of $n$ particles, because a general such state can be a \emph{superposition} of possibly exponentially many classical configurations. I.e., for a general pure quantum state one cannot assign a definite pure state to each of the constituent particles as the parts do not completely determine the whole. The mismatch here is due to \emph{quantum entanglement}, coming from the quantum correlations between the constituent particles. Thus, in order to completely describe the reduced state of a single particle we also need to keep track of the quantum correlations that develop between it and the rest of the system. This is, generically, difficult.

However, we can exploit locality structure to place bounds on how fast information can propagate through the system under natural dynamics. The physical picture to have in mind here is that of a lattice of particles connected by springs: if you wiggle one particle it should take a time proportional to the distance multiplied by a ``speed of sound'' for your disturbance to propagate through the system. This picture does remain metaphorically correct in the quantum setting and a fundamental result known as the \emph{Lieb-Robinson bound} quantifies this intuition: suppose that $H$ is the hamiltonian for a quantum spin system on a regular lattice and $A$ and $B$ are two observables acting nontrivially on two disjoint regions, then \cite{lieb:1972a} (see \cite{bratteli:1997a} for a textbook treatment) 
\begin{equation}
	\|[e^{itH}Ae^{-itH}, B]\| \le c\|A\|\|B\|e^{v|t|-\kappa L},
\end{equation}
where $c, v$, and $\kappa$ are constants, and $L$ is the distance\footnote{Measured using the natural lattice distance.} between the regions that $A$ and $B$ act on. The Lieb-Robinson bound says that regardless of the initial state of the system, if Alice implements some operation at time $T=0$ then no matter what observable Bob subsequently measures at some later time $T=t$ the amount of information that Alice can send to Bob is exponentially suppressed as the separation between them is increased \cite{bravyi:2006a}. Note, however, that as $t$ increases the bound becomes exponentially weaker. This trade-off between exponentials gives rise to an effective ``information propagation cone'': information about Alice's operation is exponentially suppressed outside of the region $\{x\,|\, d(x, A) \le \frac{v}{\kappa}|t| \}$, where $d(x, A)$ measures the distance between site $x$ and Alice's region. 

The proof of Lieb-Robinson bound has been simplified and generalised to a variety of settings beyond the original translation invariant setting  (see, e.g., \cite{nachtergaele:2010b} for a recent review): beginning with \cite{marchioro:1978a, radin:1978a} we've witnessed a sequence of results taking \cite{hastings:2004a, hastings:2005a, nachtergaele:2005a, nachtergaele:2006a, eisert:2006b} it beyond the translation-invariant setting to systems with disorder \cite{burrell:2007a}, classical systems \cite{raz:2009a, amour:2010a}, unbounded operators \cite{butta:2007a, cramer:2008a, nachtergaele:2009a, nachtergaele:2010a, premont-schwarz:2010a, premont-schwarz:2010b}, and dissipative systems \cite{poulin:2010a, nachtergaele:2011a}. Applications of the Lieb-Robinson idea have had influences further afield, e.g., to give new proof techniques for understanding the emergence of effective equations in mean-field settings \cite{erdos:2009a}.

Armed with the Lieb-Robinson bound we now have the necessary quantitative tool to understand exactly how to parametrise naturally occurring states. Intuitively, although each particle does become correlated with the others after a small time, it should only be able to develop correlations with its neighbours within a distance proportional to the Lieb-Robinson velocity. One can then keep track of only these correlations and obtain a compact description of the global state. Thus we obtain an \emph{effectively local} description of the system's state, i.e., we specify the state by recording the correlations between any given particle and only its immediate neighbours. The remarkable consequence \cite{osborne:2005d, hastings:2008a, hastings:2009b} is that, in one dimension, there is an efficient representation known as a \emph{matrix product state} (MPS) for the system's state, as long as $|t| \lesssim \log(N)$. An algorithm exploiting this phenomenon was already proposed in \cite{vidal:2003a}; the preceding results can be interpreted as saying that this algorithm (and relatives) is indeed correct, converges, and has polynomial complexity. That is, the real-time simulation problem is in $\mathsf{P}$. Further advances along these lines have shown that the simulation problem for \emph{adiabatic evolution} of \emph{gapped} quantum spin systems also lies within $\mathsf{P}$ \cite{osborne:2006a}. 

The Lieb-Robinson technology has also been put to great effect by Hastings and coworkers in the study of the equilibrium properties of quantum lattice systems with a \emph{spectral gap} between the lowest two eigenvalues. Beginning with the breakthrough result \cite{hastings:2004a}, where the the \emph{Lieb-Schultz-Mattis} conjecture was proven in higher dimensions, much insight has been gained into the general properties of gapped quantum lattice systems (see also \cite{nachtergaele:2007a}). Firstly, the presence of the gap allows one to play the Lieb-Robinson bound off against the natural decay induced by the gap to show that all spatial ground-state correlators decay exponentially with separation \cite{hastings:2004a, hastings:2004b, nachtergaele:2005a, hastings:2005b, hamza:2009a, bachmann:2011a}. This result already hints at a more powerful generalisation: perhaps the entanglement entropy, as measured by the von Neumann entropy $S(\rho_L) = -\tr(\rho_L\log(\rho_L))$, where $\rho_L$ is the reduced density operator of a region of length $L$ in the ground state of a gapped spin system, is bounded by a constant? Indeed this is the case for one-dimensional systems and such an \emph{entropy/area} law was proved in \cite{hastings:2007a}. This result can be interpreted as placing the ground-state simulation problem for gapped $1$D spin systems \emph{in} $\mathsf{NP}$, i.e., there exists a polynomial sized proof, written as an MPS, that the ground-state energy is low. This problem is not known to be in $\mathsf{P}$.

The proof technique developed in \cite{hastings:2007a} also established the key result that the ground-state of a gapped one-dimensional quantum lattice system may be efficiently approximated by an MPS. The proof is not constructive, however, and only supplies a promise that such a representation exists. It is tempting to suppose that a simple application of the variational method to MPS should find such a representation easily. However, a naive application of the variational principle to MPS runs into problems because the objective function is a nonlinear function with many local minima, and it is far from clear how to carry out the minimisation efficiently. 

It turns out, however, that a practical solution to this problem has been known for many years under the guise of the \emph{density matrix renormalisation group} (DMRG) \cite{white:1992a}. The DMRG is now understood to be a \emph{relaxation} of the variational method applied to MPS \cite{schollwock:2011a}. While the method is a heuristic, and it does suffer from local minima problems, extensive practical numerical experience has shown that these local minima do not cause any significant problems. Indeed, the DMRG is, without peer, the premiere method for the exploration of strongly correlated quantum physics in one dimension and is hard to overstate the impact of the DMRG in the field of condensed matter physics. 

The question of whether there does exist a polynomial algorithm to vary over MPS has, however, been addressed. It was recently established \cite{aharonov:2010a, schuch:2010a} that a transfer-operator type argument yields an efficient method to vary over MPS with a fixed number of variational parameters. Unfortunately, the performance of this algorithm is not comparable with the DMRG, and is unlikely to supplant the DMRG as the method of choice in the study of one-dimensional quantum systems. 

The MPS variational class was proposed and studied well before the advent of hamiltonian complexity. But with the advent of quantum information theory and hamiltonian complexity new ideas began to flow into condensed matter physics leading to the development of several new variational classes for quantum lattice systems. One important example are the \emph{projected entangled-pair states} (PEPS) \cite{verstraete:2004a}, which are a natural generalisation of MPS to the two-dimensional setting. Another example is the \emph{multiscale entanglement renormalisation ansatz} (MERA) \cite{vidal:2006a, vidal:2007a}, which is particularly well-suited to the description of low-dimensional quantum lattice systems at or near cricality. Both of these classes have been put to great effect, in combination with the variational principle, in the practical study of quantum lattice systems. 

Parallel to these theoretical developments there has been the exploration of \emph{classical} heuristics and algorithms to approximate ground-state properties of quantum spin systems. Firstly, it was realised \cite{bravyi:2006b} that one can exploit the low local dimensionality of quantum spins to efficiently decide the quantum $2$-satisfiability problem in the spin-$1/2$ case. A variation on this theme exploits possible \emph{low-degree} structure an interaction hypergraph to supply a criteria to decide whether a quantum satisfiability instance \emph{always} has a solution. This result \cite{ambainis:2010a} is known as the \emph{quantum Lovasz local lemma}. Further afield, the study of approximation algorithms and heuristics for $\mathsf{QMA}$ hard instances of the local hamiltonian problem has been incepted in \cite{bansal:2009a, gharibian:2011a}.

Recently, in another direction, Lieb-Robinson ideas and a technique known as \emph{quasi-adiabatic continuation} \cite{hastings:2005a, bachmann:2011a} have been exploited to prove several powerful results concerning the stability of topological order  \cite{bravyi:2010b, bravyi:2010c}. Quasi-adiabatic continuation has been further explored in the context of disordered systems \cite{hastings:2010b}, where the improved bound on information flow coming from the Anderson localisation phenomena may be exploited to improve many results concerning the decay of correlations in strongly interacting quantum systems. One of the most recent triumphs of these research directions has been the proof \cite{hastings:2009a} of an outstanding conjecture in mathematical physics, namely, the quantisation of the Hall conductance without averaging assumptions. 

\subsection{Guiding problem: entanglement entropy/area laws}
The entanglement entropy/area law states that for a quantum system in the ground state the entropy of any subregion $A$ should scale as the area of the boundary. This result may be regarded as an expression of the \emph{holographic principle}, namely, that the physics in the bulk of the system can be completely described by what happens at the boundary. The existing proofs of these results have been influential in hamiltonian complexity and condensed matter physics, typically leading to many new insights and results, including new efficient methods to simulate strongly correlated systems.

Quantitative results concerning the entropy/area law are, however, difficult to obtain. Indeed, apart from special cases \cite{masanes:2009a}, the only settings where the result has been proved at the level of mathematical rigour pertain to gapped one-dimensional quantum spin systems and gaussian systems (see, e.g., \cite{eisert:2010b} for a review). One generalisation that has proved particularly elusive is to gapped two-dimensional quantum spin systems. Here the existing proof ideas are extremely suggestive, yet just fail to yield the expected  law. The difficulty is probably tied, in part, to the fact that two-dimensional quantum systems can exhibit \emph{topological order}, which is difficult to capture using the existing proof techniques. It is clear that whatever method is developed to tackle this case will have important consequences not only for the simulation of two-dimensional quantum systems, but also for the study of exotic phases of matter.

\section{Conclusion}
Hamiltonian complexity, as a field, has developed rapidly in the past couple of years amassing a broad collection of foundational results exploiting a wide portfolio of techniques. Highlights include the proof that it is $\mathsf{QMA}$-complete to approximate the ground-state energy of even a one-dimensional quantum lattice system and, in the other direction, proofs that the imaginary time simulation problem is in $\mathsf{NP}$ for \emph{gapped} one-dimensional quantum lattice systems, and  the real-time simulation problem is in $\mathsf{P}$ for $|t|\sim \log(n)$. 

The task of finding quantum lattice systems which are hard to simulate appears to have reached a natural pause, of sorts. This is probably due to the difficulty of the proofs required and the fact that the existing techniques have been pretty much pushed as far as they can go. However, several simple-sounding problems remain, such as the commuting hamiltonian problem, and the search for inefficient quantum PCPs, and progress on these would open up many new directions. 

The development of efficient algorithms for the simulation of low-dimensional quantum lattice systems continues unabated. In this direction I think the study of quantum approximation algorithms initiated in \cite{gharibian:2011a} holds particular promise.  

One area that I am particularly excited about is that of the generalisation of the insights obtained in hamiltonian complexity to quantum systems with \emph{continuous} degrees of freedom, including, quantum fields; we've recently understood \cite{verstraete:2010a, osborne:2010a, haegeman:2011a} how to generalise the two main variational classes studied in the hamiltonian complexity literature, namely MPS and MERA, to the continuous setting, and it seems likely that many new results generalising those originating in the lattice setting will be possible.

\paragraph{\it Acknowledgements.} I would like to sincerely thank Sergey Bravyi, Mick Bremner, Jens Eisert, Daniel Gottesman, Matt Hastings, Bruno Nachtergaele, Barbara Terhal, and Frank Verstraete for their careful reading of preliminary drafts of this manuscript, and their many helpful comments and suggestions. This work was supported, in part, by the cluster of excellence EXC 201 ``Quantum Engineering and Space-Time Research'', by the Deutsche Forschungsgemeinschaft (DFG). Thanks go also to Il Vicino, Sante Fe, where it all got started. 
\providecommand{\bysame}{\leavevmode\hbox to3em{\hrulefill}\thinspace}
\providecommand{\MR}{\relax\ifhmode\unskip\space\fi MR }
\providecommand{\MRhref}[2]{%
  \href{http://www.ams.org/mathscinet-getitem?mr=#1}{#2}
}
\providecommand{\href}[2]{#2}

\end{document}